\documentclass[preprint,nofootinbib,amsmath,amssymb,floatfix]{revtex4}
\usepackage{graphicx}
\numberwithin{equation}{section}

\begin{document}

\title{The  Quantum Hall Effect in Graphene} 

\author{Paolo Cea$^{1,2}$}
\email[]{Paolo.Cea@ba.infn.it}

\affiliation{$^1$Dipartimento di Fisica dell'Universit\`a di Bari, I-70126 Bari, Italy \\
$^2$INFN - Sezione di Bari, I-70126 Bari, Italy }

\begin{abstract}
We investigate the quantum Hall effect in  graphene.  We argue that  in  graphene  in presence of an external magnetic field there is  dynamical generation of mass by a rearrangement of the Dirac sea. We show that the mechanism  breaks the lattice valley degeneracy  only for the $n=0$ Landau levels and leads to the new observed $\nu =  \pm 1$ quantum Hall plateaus. We suggest that our result can be tested by means of numerical simulations of planar Quantum Electro Dynamics  with dynamical fermions in an external magnetic fields on the lattice.
\end{abstract}



\maketitle

\section{\normalsize{Introduction}}
\label{introduction}
Graphene~\cite{Novoselov:2004,Novoselov:2005} is a flat monolayer of carbon atoms tightly packed in   a two dimensional honeycomb lattice consisting of two interpenetrating triangular sublattices. The structure of graphene has attracted considerable attention   from  the fact  that the low electronic excitations are described by the Dirac equation for two dimensional  massless fermions~\cite{NovoselovGeim:2005,Zhang:2005} (for  recent reviews see Refs.~\cite{Geim:2007,Katsnelson:2007,CastroNeto:2009,Kotov:2010} ).  To describe the low-energy excitations one may consider the excitations at the Fermi level in the vicinity of  two Dirac cones centered at two inequivalent corners of the Brillouin zone (called valley) $K_{\pm} \, = \, \pm \, \frac{4 \, \pi }{3 \sqrt{3} \, a_0}$, where $ a_0 \, = \, 1.42 \, 10^{-8} \, cm$ is the carbon-carbon distance. As a result, one obtains an effective Hamiltonian made of two copies of Pauli spinors $\Psi_{\pm}$ which satisfy the massless two-dimensional Dirac equation with the speed of ligth replaced by the Fermi velocity $v_F  \simeq  1.0 \, 10^8  \, cm/s$.  
In graphene, time reversal symmetry and  parity, i.e., spatial inversion symmetry with respect to the center of a hexagon, transform electronic states between valleys. Thus,
 the energy spectrum is degenerate between the different valleys since the time reversal operation connects electronic states at $K_+$ to those at $K_-$.   
 If there is dynamical generation of gap $\Delta_0$, then the energy levels of two valleys are related as  $\varepsilon_{K_+}(\Delta_0) = \varepsilon_{K_{-}} (-\Delta_0)$ due to the inversion symmetry. \\
In graphene immersed in transverse magnetic field, the relativistic massless dispersion of the electronic wave functions results in non-equidistant Landau levels (cgs units):
\begin{equation}
\label{eq1.1}
\varepsilon_n =  sign(n) \sqrt{2 |n| \hslash \frac{ v_F^2}{c} eH} \, , \,  n \, = \, 0 \, , \, \pm \, 1 \, , \, \pm \, 2  \; ...
\end{equation}
where $eH>0$, $e$ being the elementary charge. The presence of anomalous Landau levels at zero energy, $\varepsilon_0=0$, leads to half-integer quantum Hall 
effect~\cite{NovoselovGeim:2005,Zhang:2005,NovoselovMcCann:2006}:
\begin{equation}
\label{eq1.2}
R_{xy}^{-1}  = \pm  g_s  \frac{e^2}{2 \pi  \hslash} (N + \frac{1}{2})   \, , \,  N \, = \, 0 \, , \, 1 \, , \, 2  \; ...
\end{equation}
where $R_{xy}$ is the Hall resistance, $g_s=4$ accounting for spin and valley degenerancy, and $\pm$ stands for electrons and holes, respectively. This quantization condition corresponds to quantized filling factor:
\begin{equation}
\label{eq1.3}
\nu   = \pm  g_s (N + \frac{1}{2})  = \pm 2 \, , \,  \pm 6 \, , \, \pm 10  \,\,   ...
\end{equation}
Recently,  experimental investigations of quantum Hall effect in graphene in very strong magnetic field $H \gtrsim 20 \, T$ ($1 \, T = 10^4 \, gauss$) have revealed new quantum Hall states corresponding to filling factor $\nu   = 0 \, , \, \pm  1 \, , \,  \pm 4$~\cite{Zhang:2006,Jiang:2007}. It is wiedely believed that the new plateaus at $\nu   = 0 \,  \, , \,  \pm 4$ can be explained by Zeeman spin splitting. On the other hand the  $\nu   =  \pm 1$ plateaus are associated with the spontaneuos breaking of the valley symmetry in the $n=0$ Landau levels. Indeed, these states are naturally explained if there is dynamical generation of a gap $\Delta_0$ .  Moreover, it turns out that $\Delta_0(H) \sim \sqrt{H}$~\cite{Jiang:2007}.  \\
This peculiar dependence on the magnetic field suggests that spontaneous generation of the gap is driven
by the electron-electron Coulomb interaction~\cite{Khveshchenko:2001,Gusynin:2006,Herbut:2007,Gusynin:2007,Kotov:2010}.
In this picture, the dynamical gap is expected to be:
\begin{equation}
\label{eq1.6}
\Delta_0(H) \, \simeq \, \frac{e^2}{\varepsilon} \, \sqrt{\frac{eH}{\hslash c}} \,   \simeq \,  163 \; ^0K \, k_B \; \sqrt{H(T)} 
\end{equation}
assuming the dielectric constant $ \varepsilon \simeq 4$~\cite{Jiang:2007}. The aim of this paper is to discuss a different mechanism which is able to spontaneuosly break  the valley symmetry in presence of an external magnetic field. Indeed, we show  that by a rearrangement of the Dirac sea 
the vacuum energy density  is lowered by inducing a dynamical gap $\Delta_0 \sim \sqrt{H}$.  \\
The plan of the paper is as follows. In Sect.~\ref{gap} we present the dynamical generation of a gap  by a rearrangement of the Dirac sea in presence of 
an external magnetic field.  In Sect.~\ref{condensate} we discuss the fermion condensate in the one loop approximation. Finally, our conclusions are relegated in 
Sect.~\ref{conclusion}.
\section{\normalsize{ Dynamical generation of the mass gap}}
\label{gap}

In this Section we discuss the mechanism which is able to spontaneously break  the valley symmetry 
in presence of an external magnetic field by a rearrangement of the Dirac sea. In particular, we show that 
the vacuum energy density  is lowered by inducing a dynamical gap $\Delta_0 \sim \sqrt{H}$.  \\
\begin{figure}
\includegraphics[width=0.9\textwidth,clip]{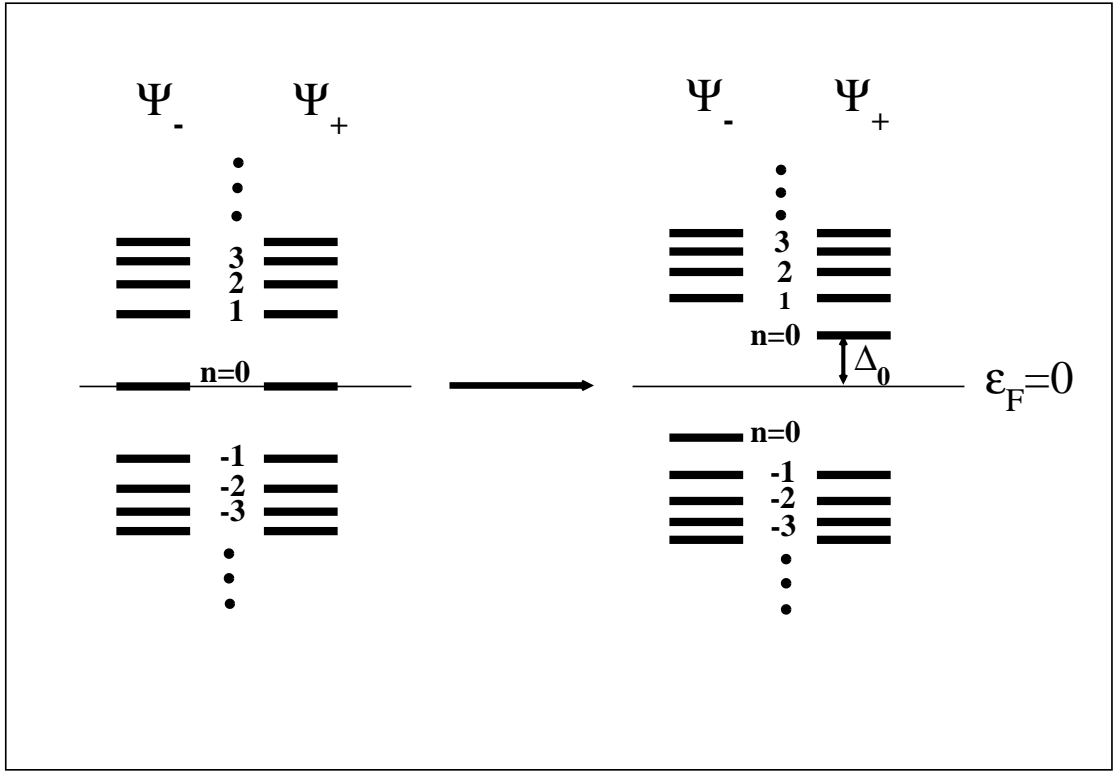}
\caption{\label{fig-1}
Schematic spectrum of Landau levels of graphene in applied magnetic field (left). Landau levels with  dynamical generation of a gap $\Delta_0$ (right).  
The Fermi level is at $\varepsilon_F =  0$.}
\end{figure}
We said that  the low-energy excitations in graphene are described by two Pauli spinors $\Psi_{\pm}$  which satisfy the massless two-dimensional Dirac equation.
When  graphene is immersed in a transverse magnetic field, the relativistic massless dispersion of the electronic wave functions results in non-equidistant Landau levels Eq.~(\ref{eq1.1})
for both  fields  $\Psi_{\pm}$ (see Fig.~\ref{fig-1}, left).   If there is dynamical generation of gap $\Delta_0$, since  the energy levels of two valleys are related as  $\varepsilon_{K_+}(\Delta_0) = \varepsilon_{K_{-}} (-\Delta_0)$ due to the inversion symmetry, the $\Psi_{\pm}$ are no longer degenerate. Indeed, it is easy to find the following spectrum (see Fig.~\ref{fig-1}, right):
\begin{equation}
\label{eqS1.2}
\varepsilon_{+,0} = \Delta_0 \;\; , \;\;\; \varepsilon_{+,n} =  \text{sign}(n) \sqrt{2 |n| \hslash \frac{ v_F^2}{c} eH+\Delta_0^2} \, , 
\;\;\;\;\;  n \, = \,  \pm \, 1 \, , \, \pm \, 2  \; ...\;,
\end{equation}
\begin{equation}
\label{eqS1.3}
\varepsilon_{-,0} =  - \Delta_0 \;  , \;  \varepsilon_{-,n} =  \text{sign}(n) \sqrt{2 |n| \hslash \frac{ v_F^2}{c} eH+\Delta_0^2} \, , 
\;\;\;\;\;  n \, = \,  \pm \, 1 \, , \, \pm \, 2  \; ...\;,
\end{equation}
for the  $\Psi_{+}$ and  $\Psi_{-}$ fields respectively. \\
Let us consider the vacuum energy density ${{\cal E}}$ obtained by filling the Dirac sea. For massless fields, according to Eq.~(\ref{eq1.1}) we obtain:
\begin{equation}
\label{eqS1.4}
{{\cal E}}(0)  \, = \, - 2 \,  \frac {eH} {2 \pi \hslash c}  \sum_{n=1}^{\infty}  \sqrt{2 n \hslash \frac{ v_F^2}{c} eH} \; ,
\end{equation}
where the factor in front of the sum takes care of the degeneracy of the  Landau levels. In presence of a gap $\Delta_0$, using Eqs.~(\ref{eqS1.2}) and (\ref{eqS1.3}), we find:
\begin{equation}
\label{eqS1.5}
{{\cal E}}(\Delta_0)  \, = \,  \frac {eH} {2 \pi \hslash c}   \left( - \Delta_0 \, - 2 \,  \sum_{n=1}^{\infty}  \sqrt{2 n \hslash \frac{ v_F^2}{c} eH+\Delta_0^2} \,  \right )
\end{equation}
So that we get:
\begin{eqnarray}
\label{eqS1.6}
&& \Delta{{\cal E}}  \equiv  {{\cal E}}(\Delta_0) - {{\cal E}}(0)   \, = \,  
\nonumber \\
&& \frac {eH} {2 \pi \hslash c}   \left( - \Delta_0 \, - 2 \,  \sum_{n=1}^{\infty}  \sqrt{2 n \hslash \frac{ v_F^2}{c} eH+\Delta_0^2} \,  
+ 2 \,    \sum_{n=1}^{\infty}  \sqrt{2 n \hslash \frac{ v_F^2}{c} eH} \, \right ) .
\end{eqnarray} 
The sums in  Eq.~(\ref{eqS1.6}) are divergent. To evaluate $ \Delta{{\cal E}}$ we adopt the Schwinger's proper time regularization~\cite{cea:1985}. 
To this end,  suffice it to use the integral  representation :
\begin{equation}
\label{eqS1.7}
\sqrt{a} = - \int_0^{\infty} \frac {d s} {\sqrt{ \pi s}} \; \frac {d} {d s} \; 
 e^{- a s}  \; .
\end{equation}
A straightforward calculation gives:
\begin{equation}
\label{eqS1.8}
\Delta \varepsilon \equiv \frac{\Delta{{\cal E}}}{ \frac {eH} {2 \pi \hslash c}} \, = \,  \sqrt{ \hslash \frac{ v_F^2}{c} eH}  \;  \left [\;  - \lambda \, + \, g(\lambda)  \; \right  ]
\; \; \; , \; \; \; \lambda \, = \, \frac{\Delta_0}{ \sqrt{ \hslash \frac{ v_F^2}{c} eH}} \; ,
\end{equation}
where:
\begin{equation}
\label{eqS1.9}
 g(\lambda) \, = \, 2 \,  \int_0^{\infty} \frac {d x} {\sqrt{ \pi x}} \; \frac {d} {d x} \;  \left [
 \left ( e^{- \lambda^2 x} - 1\right )  \; \frac{e^{-2 x}}{1-e^{-2 x}} \; \right ] \; .
\end{equation}
\begin{figure}
\includegraphics[width=0.9\textwidth,clip]{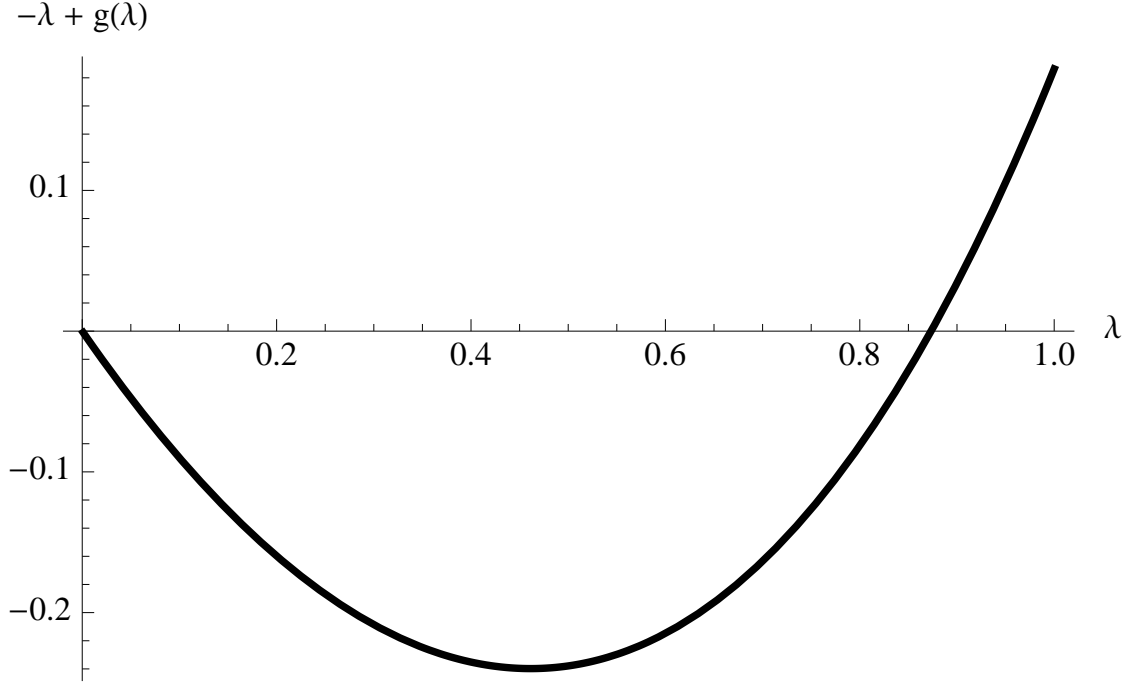}
\caption{\label{fig-2}
Plot of the function  $- \lambda \, + \, g(\lambda)$ versus $\lambda$.}
\end{figure}

In Fig.~~\ref{fig-2} we display the function $- \lambda \, + \, g(\lambda)$. It is evident that there is a negative minimum for $\lambda = \overline \lambda \simeq 0.45$. 
Thus, we see that it is energetically favourable to induce a dynamical gap:
\begin{equation}
\label{eqS1.10}
 \Delta_0 \; = \overline \lambda \;  \sqrt{ \hslash \frac{ v_F^2}{c} eH} \; \simeq \;   134  \; {\rm ^0K} \times k_B \; \sqrt{H(T)} ,
\end{equation}
which is comparable to  the estimate of the  gap  driven by the electron-electron Coulomb interaction, Eq.~(\ref{eq1.6}) .
\section{\normalsize{Chiral Condensate }}
\label{condensate}
Recently, it has been obseved that  the low-energy dynamics of graphene can be accounted for by $N_f=2$ massless
 Dirac fields~\cite{Drut:2009}. These are descibed nonperturbatively by the lattice Euclidean action using  staggered fermion fields~\cite{Hands:2002}.
 As a consequence, we see that our mechanism for the dynamical generation of gap by a rearrangement of the Dirac sea  
 can be tested by means of numerical simulations of planar Quantum Electro Dynamics  with dynamical fermions in an external magnetic field on the lattice.
In particular, the dynamical generation of the gap can be easily tested on the lattice by measuring the fermion  condensate in the chiral limit. To see this, following
 Ref.~\cite{cea:2000}, we discuss relativistic fermions coupled with electromagnetic field in (2+1) dimensions. For reader convenience we do not follow the high-energy natural units where $\hslash = c = 1$, but we employ cgs units. \\
A spinorial representation in three dimensions is provided by two-component  Dirac spinors. The fundamental representation of the Clifford algebra is given 
by $2 \times 2$ matrices which can be constructed from the Pauli matrices:
\begin{equation}
\label{eqS2.1}
\gamma^0=\sigma_3, \; \; \gamma^1 =i \sigma_1, \; \; \gamma^2=i \sigma_2 \; .
\end{equation}
Note that in this representation the mass term  $m c^2 {\bar {\Psi}} \Psi$ is 
odd under both parity and time-reversal transformations (for a review, see Ref.~\cite {Jackiw:1982}).  We are interested in the chiral condensate which is defined as:
\begin{equation}
\label{eqS2.2}
\langle\overline\Psi\Psi\rangle = \frac{1}{V}  \; \int d^2 x  \;  \langle 0 | {\bar {\Psi}}(x)  \Psi(x) | 0  \rangle  \; ,
\end{equation}
where $V$ is the spatial volume and $ | 0  \rangle $ is the ground state (vacuum).  In presence of an external constant magnetic field, it is easy to evaluate the chiral condensate
in the one-loop approximation~\cite{cea:2000}:
\begin{equation}
\label{eqS2.3}
\langle \overline\Psi\Psi \rangle \;  = \frac{ \hslash  c  eH} {4 \pi}  - \, m c^2 \, \frac { \hslash  c  eH} {2 \pi} 
\sum_{n=1}^{\infty} \frac {1} {\sqrt{2 n \hslash c eH+m^2 c^4}} \; .
\end{equation}
It is worthwhile to stress that the first term in Eq.~(\ref{eqS2.3}), due to the $n=0$ Landau levels, is independent on the mass and it is related to the parity anomaly. In fact, we see
that the fermion condensate of the zero-mass theory in the presence of an external constant magnetic field corresponds to half-filled zero modes, which in turns leads to the observed half-integer quantum Hall effect in graphene. \\   
In three dimensions we have a realization of the  Dirac algebra which is different from the $2 \times 2$ representation Eq.~(\ref{eqS2.1}). 
In this representation the Dirac fermions are four-component spinor and the  three $4 \times 4$ $\gamma$-matrices can be taken to be~\cite{Appelquist:1986}:
\begin{equation}
\label{eqS2.4}
\gamma^0= \left( \begin{array}{clcr}
                   \sigma_3  & \; \; \; 0 \\
                   0 & - \sigma_3  
                   \end{array} \right) \;\;\;
\gamma^1= \left( \begin{array}{clcr}
                    i \sigma_1  & \; \; \; 0 \\
                   0 & -i \sigma_1  
                   \end{array} \right) \;\;\;
\gamma^2= \left( \begin{array}{clcr}
                    i \sigma_2  & \; \; \; 0 \\
                   0 & -i \sigma_2  
                   \end{array} \right) \;\;\; .
\end{equation}                    
The representation given by  Eq.~(\ref{eqS2.4}) corresponds  to a reducible representation of  the Dirac algebra.
In this representation the fermionic mass term  $m {\bar {\Psi}} \Psi$  is parity conserving. Indeed,
comparing  Eq.~(\ref{eqS2.1}) with Eq.~(\ref{eqS2.4}) we see that a four-component fermion with
mass $m$ corresponds to two two-component fermions with mass $m$ and $-m$ respectively.  So that, if we write
 $ \Psi=\left(\begin{array}{c}\Psi_+  \\ \Psi_- \end{array} \right)$, we readily get:
\begin{equation}
\label{eqS2.5}
\langle \overline\Psi \Psi \rangle \; = \; \langle \overline\Psi_+ \Psi_+ \, - \, \overline\Psi_- \Psi_- \rangle \; .
\end{equation}                    
Since under parity transformation $\Psi_+ \rightarrow \Psi_-$ and $\Psi_- \rightarrow \Psi_+$, we see that the chiral condensate Eq.~(\ref{eqS2.5}) is
parity invariant.   In presence of an external constant magnetic field, we may  easily  evaluate the chiral condensate in the one-loop approximation
by using Eqs.~(\ref{eqS2.3})  and (\ref{eqS2.5}):
\begin{equation}
\label{eqS2.6}
\langle \overline\Psi \Psi \rangle \; = \; - \, 2 \, |m| c^2 \, \frac { \hslash  c  eH} {2 \pi} 
\sum_{n=1}^{\infty} \frac {1} {\sqrt{2 n \hslash c eH+m^2 c^4}} \; .
\end{equation}                    
Using the following identity:
\begin{equation}
\label{eqS2.7}
\frac {1} {\sqrt{a}} \; = \; \int_0^{\infty}  \, \frac {ds} {\sqrt{\pi s}} \; e^{-as} \; ,
\end{equation}
we recast Eq.~(\ref{eqS2.6}) into:
\begin{eqnarray}
\label{eqS2.8}
&&  \langle \overline\Psi \Psi \rangle \; = \;  - \, 2 \, |m| c^2 \, \frac { \hslash  c  eH} {2 \pi}  \frac{1}{\sqrt{2  \hslash c eH}}
 \int_0^{\infty} \frac {d x} {\sqrt{ \pi x}} \; 
 e^{- \alpha^2 x}   \; \frac{e^{- x}}{1-e^{- x}} \; \; , 
 \nonumber \\
 && \; \; \; \;  \; \; \alpha \, = \, \frac{|m| c^2}{\sqrt{2  \hslash c eH}} \; .
\end{eqnarray}                
Now, we may regularize the integral by analytic continuation. To do this, we note that~\cite{Gradshteyn:1980}:
\begin{equation}
\label{eqS2.9}
\int_0^{\infty}  \, d x  \;  \frac{x^{\nu-1} \, e^{-\mu x}}{1-e^{-\beta x}} \; \; = \; \; \frac{1}{\beta^{\mu}} \; \Gamma(\nu) \; 
 \zeta(\nu, \frac{\mu}{\beta}) \; ,
\end{equation}                    
where  $\Gamma(x)$ is the Euler gamma function and  $\zeta(x,y)$ is the generalized Riemann Zeta function.
Combining Eqs.~(\ref{eqS2.8}) and (\ref{eqS2.9}) we finally obtain:
\begin{equation}
\label{eqS2.10}
\langle \overline\Psi \Psi \rangle \; = \; - \, 2 \, |m| c^2 \, \frac { \hslash  c  eH} {2 \pi}  \frac{1}{\sqrt{2  \hslash c eH}}
\frac{\Gamma(\frac{1}{2})} {\sqrt{ \pi}} \;  \zeta(\frac{1}{2}, 1+\alpha^2) \; , \; \alpha \, = \, \frac{|m| c^2}{\sqrt{2  \hslash c eH}} \; .
\end{equation}
Our results can be applied to graphene if the Coulomb interaction can be neglected. In that case, observing  that $\Delta_0 = |m| c^2$ 
we readly obtain from Eq.~(\ref{eqS2.10}):
\begin{eqnarray}
\label{eqS2.11}
&& \langle \overline\Psi \Psi \rangle \; = \; - \, 2 \,\Delta_0 \, \frac { \hslash  c  eH} {2 \pi}  \frac{1}{\sqrt{2  \hslash \frac{v_F^2}{c} eH}}
\frac{\Gamma(\frac{1}{2})} {\sqrt{ \pi}} \;  \zeta(\frac{1}{2}, 1+\alpha^2) \; , 
\nonumber \\
&& \; \; \; \; \; \; \;  \alpha \; = \;   \frac{\Delta_0}{\sqrt{2  \hslash  \frac{v_F^2}{c} eH}} \;  \; .
\end{eqnarray} 
For small gap, we may  expand to the first order in $\Delta_0$. Using
$\Gamma(\frac{1}{2}) = \sqrt{ \pi}$ and $\zeta(x,1) =  \zeta(x)$, we get:
\begin{equation}
\label{eqS2.12}
\langle \overline\Psi \Psi \rangle \; = \; - \, 2 \,\Delta_0 \, \frac { \hslash  c  eH} {2 \pi}  \frac{1}{\sqrt{2  \hslash \frac{v_F^2}{c}  eH}}
 \;  \zeta(\frac{1}{2}) \; .
\end{equation}
This last equation can be used to extract the gap $\Delta_0$ from non perturbative measurements of the fermion condensate on the lattice. 
Indeed, preliminary   Monte Carlo simulations ~\cite{cea:2011} of  planar Quantum Electro Dynamics  with two degenerate staggered fermions 
on the lattice suggest that  external magnetic fields give rise to the spontaneous breaking of the chiral symmetry in the continuum limit. 
\section{\normalsize{Conclusions}}
\label{conclusion}
We have discussed the  spontaneous generation of the gap in graphene induced by a rearrangement of the Dirac sea.  We find that 
the gap $\Delta_0(H)  \sim \sqrt{H}$, and it is  comparable to  the  gap  driven by the electron-electron Coulomb interaction.
We argued that our  dynamical generation of the gap can be easily tested on the lattice by measuring the fermion  condensate in the chiral limit.

\end{document}